\documentclass[journal]{vgtc}                     

\onlineid{0}

\vgtccategory{Research}

\title{Design Knowledge in Data Visualization: Mapping the Epistemic Landscape}

\author{%
  \authororcid{Paul C. Parsons}{0000-0002-4179-9686},
  \authororcid{Colin M. Gray}{0000-0002-7307-1550},
  \authororcid{Ali Baigelenov}{0009-0003-6491-1874}
}

\authorfooter{
  \item
  	Paul Parsons is with Purdue University.
  	E-mail: parsonsp@purdue.edu.
  \item
  	Colin Gray is with Indiana University.
  	E-mail: comgray@iu.edu.

  \item Ali Baigelenov is with Purdue University and SDU University.
  	E-mail: ali.baigelenov@sdu.edu.kz.
}

\abstract{%
Data visualization research has developed many influential forms of design knowledge, including perceptual principles, design guidelines, process models, and formalized representations of design constraints. These contributions have been effective at articulating explicit, portable, and codified forms of knowledge. Yet the broader landscape on which visualization design depends remains less clearly articulated, especially with respect to intermediate-level knowledge, precedents, tacit repertoires, and situated forms of knowing. In this paper, we draw on design theory to map this broader landscape of design knowledge in data visualization. Through this lens, we show how visualization research has built substantial strengths in some regions while leaving others comparatively underarticulated. We further argue that visualization design depends not only on knowledge artifacts such as theories, guidelines, and patterns, but also on knowledge-in-use---the situated interpretation, adaptation, and coordination of multiple forms of knowing in concrete design situations. This broader account has implications for how the field conceptualizes design expertise, evaluates and develops scholarly contributions, and approaches AI-assisted design. Rather than treating visualization design as either fully formalizable or wholly resistant to computational support, we argue for a differentiated view in which computational systems can support some forms of design knowing, while others remain inseparable from human judgment, contextual interpretation, and the ongoing reorganization of design work in practice.
}


\graphicspath{{figs/}{figures/}{pictures/}{images/}{./}} 

\usepackage{tabu}                      
\usepackage{booktabs}                  
\usepackage{lipsum}                    
\usepackage{mwe}                       
\usepackage{ccicons}                   

\usepackage{tabularx}
\usepackage{array}

\usepackage{mathptmx}                  

\begin{document}


\firstsection{Introduction}

\maketitle

Data visualization research has developed a substantial body of knowledge intended to support design. Across the literature, researchers have proposed reusable design resources such as patterns \cite{Heer2006, sedig_design_2016, elmqvist_patterns_2015, Elmqvist2011}, guidelines \cite{choi_toward_2021}, design spaces \cite{Card1997, schulz_design_2011}, principles \cite{ware_information_2019}, heuristics \cite{zuk_heuristics_2006, forsell_heuristic_2010}, and formal representations of visualization constraints \cite{schmidt_visual_2024}. Visualization researchers have also developed influential ways of conceptualizing design itself, including process models and structured approaches to designing \cite{sedlmair_design_2012, munzner_nested_2009, parsons_beyond_2026, mckenna_design_2014, meyer_nested_2015}. More recently, design knowledge has increasingly been treated as something that can be operationalized in recommendation systems, knowledge bases, and LLM-based design support \cite{moritz_formalizing_2019, schmidt_visual_2024, kim_data_2026, wang_dracogpt_2025}. These efforts reflect a longstanding concern with how visualization design can be understood, guided, and improved.

Despite this rich body of work, design knowledge in data visualization has more often appeared indirectly through specific contribution types than as an explicit conceptual topic in its own right. Recent work on visualization guidelines and design feedback suggests that practitioner-facing knowledge remains distributed across many sources, unevenly structured, and variably connected to empirical research and real-world use \cite{choi_toward_2021, kim_understanding_2026, kim_how_2025, otto_visualization_2024}. Visualization scholarship has also increasingly questioned the field's underlying epistemic assumptions more directly, including positivist orientations in design studies \cite{meyer_criteria_2020} and broader commitments to values such as universality, objectivity, and efficiency \cite{parsons_beyond_2026, saharan_critical_2026}. These developments suggest that the field is well positioned for a more explicit articulation of visualization design knowledge.

Design theory provides a useful vocabulary for articulating this broader landscape more explicitly. Across several design disciplines, scholars have argued that design depends not only on general theories or explicit guidelines, but also on intermediate-level knowledge \cite{Lowgren2013, dalsgaard_between_2014}, precedents \cite{boling_nature_2021}, strong concepts \cite{hook_strong_2012}, repertoires, and tacit forms of knowing developed in practice \cite{archer_design_1979, cross_designerly_1982, schon_reflective_1983, buchanan_wicked_1992, nelson_design_2012, zimmerman_designing_2009}. These perspectives emerged in part as a response to rationalist views that cast design primarily as the application of general, codified knowledge to instrumental problems. Instead, design scholars have emphasized that designing involves forms of interpretation, framing, and judgment that move across multiple levels of abstraction and are not exhausted by what can be fully articulated or formalized \cite{schon_reflective_1983, buchanan_wicked_1992, nelson_design_2012, Lowgren2013}.

This broader framing is especially salient in visualization because the field has been particularly successful at producing codified and portable forms of knowledge, including empirical guidelines, formalized constraint systems, and data-driven approaches for learning or updating design knowledge bases \cite{choi_toward_2021, schmidt_visual_2024, kim_data_2026}. Automated and AI-assisted design systems make the stakes of this issue especially clear, because they necessarily rely on assumptions about which aspects of visualization design can be represented computationally and which remain dependent on human interpretation and judgment \cite{parsons_design_2023}. Recent work on guideline use, design feedback, and AI-assisted design highlights both the promise and the limits of computational support, especially where design depends on examples, contextual interpretation, and experiential knowledge that are not easily captured in codified form \cite{kim_understanding_2026, kim_how_2025, wang_dracogpt_2025}. The issue is not simply which functions should be allocated to humans or computational systems, but how formalizing some forms of design knowledge reshapes the work of design itself---including what becomes visible, what is backgrounded, and what kinds of judgment and coordination remain necessary \cite{dekker_maba-maba_2002}.

By \emph{design knowledge}, we do not mean all knowledge that may be relevant to a visualization project. Rather, we use the term to refer to knowledge that informs, structures, or supports visualization design action, judgment, and the development of design outcomes. In this sense, design knowledge may appear in multiple forms, including empirical findings, guidelines, patterns, precedents, and situated repertoires of practice. This broader usage is consistent with prior design scholarship that treats design knowledge as plural in form rather than reducible to a single type of explicit theory or rule \cite{dong_structuring_2014, boling_knowledge_2025}.

In this paper, we draw on design theory to map a broader epistemic landscape of design knowledge in data visualization. We identify four important regions of this landscape: formal, abstract, and highly codified knowledge; intermediate-level knowledge; precedents and repertoires; and tacit and situated knowing. This conceptual lens helps reconsider the field's existing strengths and asymmetries, showing where visualization research has built especially strong knowledge infrastructures and where important forms of design knowledge remain more fragmented or underarticulated. It also clarifies that visualization design depends not only on static knowledge artifacts such as theories, principles, and patterns, but also on \emph{knowledge-in-use}: the situated mobilization, interpretation, and coordination of multiple forms of knowing in practice. We then discuss implications for research, education, and AI-assisted design. Our goal is not to displace existing forms of visualization knowledge, but to situate them within a more expansive account of how design knowledge works in the field.

\section{Background}

\subsection{Design Knowledge in Data Visualization}

Design knowledge is already an active concern in visualization scholarship, even if it is more often addressed through specific contribution types than as an explicit conceptual topic. Work on patterns, guidelines, recommendation systems, and AI-assisted design contributes to this broader terrain \cite{Heer2006, choi_toward_2021, kim_understanding_2026, moritz_formalizing_2019, schmidt_visual_2024, kim_data_2026, wang_dracogpt_2025, kim_how_2025}. Recent work on the research–practice gap in visualization design guidelines is instructive here. Kim et al. \cite{kim_understanding_2026} show that practitioner-facing guideline knowledge is dispersed across blogs, books, websites, design systems, and community resources, and that its relationship to empirical research is often uneven or unclear. Related work has also argued that visualization needs stronger accounts of how knowledge is transferred through process artifacts and across organizational settings \cite{otto_visualization_2024}. These efforts indicate that the field has developed many knowledge artifacts, but that the broader relationships among these artifacts—their levels of abstraction, degrees of codifiability, and connections to practice—remain less fully articulated.

Recent work has also begun to interrogate the assumptions that shape how knowledge is produced and valued. Meyer and Dykes \cite{meyer_criteria_2020} argue for an interpretivist perspective on visualization design studies, emphasizing knowledge as multiple, subjective, and socially constructed rather than solely objective and generalizable. More recently, Saharan et al. \cite{saharan_critical_2026} identify universality, objectivity, and efficiency as recurring values in normative visualization discourse and call for a more pluralistic value landscape to guide future tools, guidelines, and research practices. Related work has likewise questioned assumptions of neutrality, universality, and transparency from feminist and humanistic perspectives \cite{dignazio_data_2020, drucker_information_2017, akbaba_entanglements_2025}. These contributions do not directly articulate a framework for design knowledge in visualization, but they reinforce the need for broader reflection on the epistemic assumptions that shape what kinds of knowledge the field tends to produce and privilege.

\subsection{Design Knowledge in Design Theory}

Across several design disciplines, scholars have argued that design knowledge cannot be reduced to a single form or understood solely as abstract theory applied to local problems. Instead, design depends on multiple forms of knowing that operate at different levels of abstraction and in different relationships to practice \cite{dong_structuring_2014, nelson_design_2012}. Early accounts of designerly ways of knowing emphasized that design constitutes a distinct mode of inquiry and action, not merely an applied extension of science or the humanities \cite{archer_design_1979, cross_designerly_1982}. Related philosophical work on practice and personal knowledge likewise emphasized experience, judgment, skillful integration, and responsiveness to particular situations \cite{dunne_back_1997, polanyi_personal_1958}.

One influential framing comes from Nelson and Stolterman, who distinguish between \emph{universals} and \emph{ultimate particulars} \cite{nelson_design_2012}. In their account, scientific inquiry tends toward universal principles on the side of what is true, whereas design culminates in the production of an ultimate particular: a concrete, singular outcome brought into existence in the real world through intention and judgment. This distinction suggests that design knowledge cannot be understood only in terms of abstract truths, general laws, or formal rules. Such knowledge may orient action, but design remains accountable to the creation of a particular artifact, system, or intervention under particular conditions. Abstract principles and general laws therefore leave room for interpretation, prioritization, and judgment in the concrete case.

L\"owgren and colleagues extend this picture by emphasizing a middle layer of \emph{intermediate-level knowledge} between universal theory and particular artifacts \cite{hook_strong_2012, Lowgren2013}. This middle territory includes forms such as guidelines, heuristics, patterns, methods and tools, strong concepts, experiential qualities, and annotated portfolios—i.e., constructs that are more general than individual cases but less abstract than theory. Such knowledge is often generative rather than prescriptive. Rather than dictating fixed solutions, it helps designers see possibilities, structure situations, and orient action across families of design problems \cite{dalsgaard_between_2014, zimmerman_designing_2009}. This perspective is especially useful for visualization because many familiar contribution types—including guidelines, patterns, process models, design studies, and exemplars—appear to occupy this intermediate terrain.

Design theory has also challenged rationalist views that cast professional practice primarily as the application of general, codified knowledge to instrumental problems. Sch\"on famously described this view as \emph{Technical Rationality}, arguing that it fails to account for how practitioners actually work in situations characterized by uncertainty, uniqueness, and value conflict \cite{schon_reflective_1983}. Related critiques in design scholarship similarly argue that designing involves framing, interpretation, and judgment, rather than the straightforward application of rules \cite{buchanan_wicked_1992, dorst_comparing_1995, lowgren_thoughtful_2004}. Dunne’s critique of ``the lure of technique’’ sharpens this point by showing how practice is distorted when technique is treated as though it could substitute for practical judgment \cite{dunne_back_1997}. This broader view provides the basis for the landscape developed in the next section---one in which design knowledge may be highly codified or only partly articulable, broadly portable or strongly situated, and embodied not only in theories and rules but also in precedents, repertoires, and situated action.

\subsection{AI, Automation, and the Stakes of Design Knowledge}

These issues become especially important as visualization research increasingly explores automation and AI-assisted design. Systems such as Draco operationalize design knowledge as formal constraints, rules, and learned preferences over visualization specifications \cite{moritz_formalizing_2019, schmidt_visual_2024}. More recent work has sought to expand such knowledge bases through data augmentation \cite{kim_data_2026}, to probe whether large language models have internalized useful visualization design preferences \cite{wang_dracogpt_2025}, and to examine AI more directly as a source of design feedback and guidance for visualization creators \cite{kim_how_2025}. These lines of work are valuable, but they also make assumptions about what kinds of design knowledge can be represented, learned, and applied computationally.

Prior work has also suggested that automated visualization design systems often serve at least two distinct aims: assisting analysts who may lack visualization design expertise by recommending plausible charts or encodings, and assisting designers by representing and applying codified design knowledge more systematically in the design process \cite{parsons_design_2023}. These aims overlap, but they place different demands on what must be represented computationally. In the former case, systems may only need to identify plausible mappings or chart types. In the latter, they must encode design knowledge in forms that can support more deliberate design reasoning.

These systems also inherit broader normative assumptions from the field. As Saharan et al.\ argue, much of normative visualization has historically been shaped by values such as universality, objectivity, and efficiency \cite{saharan_critical_2026}. When design knowledge is formalized for automation, these values may become embedded not only in guidelines and evaluation criteria, but also in the computational logic of recommendation and generation systems. This helps clarify why automation debates in visualization are ultimately debates about design knowledge. Some aspects of design lend themselves to formalization, especially those involving explicit constraints, recurring heuristics, or search through spaces of known solutions. But real design situations are often more expansive, as designers do not simply search for solutions; rather, they frame what the problem is, what counts as useful, and which considerations matter in a particular context \cite{parsons_design_2023}.

Recent visualization scholarship already points to both the promise and the limits of these approaches. Work on guideline translation suggests that empirical findings do not automatically become usable design guidance in practice, and that practitioners often depend on contextual examples, interpretation, and accumulated experience alongside more formal guidance \cite{kim_understanding_2026}. Research on LLM-based design support similarly suggests that such systems can provide broad and often useful advice, while still struggling with nuanced contextual understanding and, in some cases, diverging from empirically derived best practices \cite{wang_dracogpt_2025, kim_how_2025}. The issue, then, is not whether visualization design can be automated in general, but which forms of design knowledge are amenable to formalization, retrieval, generation, or evaluation, and which remain inseparable from human judgment, interpretation, and the framing of design situations.

\section{The Epistemic Landscape of Design Knowledge in Data Visualization}

The preceding discussion suggests that design knowledge in visualization is not a single, uniform entity, but a broader epistemic landscape spanning multiple forms of abstraction and multiple relations to practice. The regions of this landscape are not intended as rigid or mutually exclusive knowledge types, nor as an exhaustive taxonomy of all possible forms of design knowledge. Rather, they are analytic regions chosen to foreground recurring distinctions in design theory concerning abstraction, codifiability, relation to artifacts, and proximity to situated practice. In this sense, they mark recurring zones of emphasis in how design knowledge is articulated, circulated, and enacted. The four regions traced here were selected because they help make visible several major contrasts that are central to our argument, especially between highly codified knowledge artifacts, practice-linked abstractions, example-based repertoires, and forms of knowing that remain inseparable from use in context. Other categorizations are certainly possible, and the boundaries among these regions are often porous in practice. What follows is a conceptual map of four especially important regions of this landscape: formal, abstract, and highly codified knowledge; intermediate-level knowledge; precedents and repertoires; and tacit and situated knowing.

\subsection{Formal, Abstract, and Highly Codified Knowledge}

At one end of the epistemic landscape of design knowledge in data visualization are forms of knowledge that are relatively abstract, highly codified, and intended to travel across contexts with limited reinterpretation. By \emph{abstract}, we do not mean wholly detached from context, but rather articulated at a level meant to extend beyond a single case by expressing more general constructs, relations, or regularities. By \emph{context}, we mean the concrete task, domain, organizational setting, and design situation in which knowledge is produced and used. These forms include formalized empirical knowledge about perception, cognition, and task performance, taxonomies and typologies that organize design spaces, framework-level theoretical contributions, formal models of design and evaluation, optimization-oriented process accounts, and computational representations of design knowledge in recommendation systems and knowledge bases. While these forms differ in scope and purpose, they share a common orientation toward articulating design-relevant knowledge in ways that are explicit, stable, and broadly reusable.

One longstanding source of such knowledge in visualization is the accumulation of empirical findings about perception, cognition, and task performance. Controlled studies have produced widely used abstractions about visual encoding effectiveness, chart interpretation, and design tradeoffs that have informed the field for decades \cite{Cleveland1984, heer_crowdsourcing_2010, franconeri_science_2021, Munzner2014}. In this region of the landscape, such knowledge is valued for its explicitness, portability, and potential for cumulative refinement. Its authority often derives from systematic empirical grounding, and its usefulness lies in its offering relatively stable guidance that can travel beyond the original study setting, even though its relevance and force may still vary with task, domain, and design context.

A second important form is classificatory knowledge: taxonomies, typologies, and reference models that organize tasks, techniques, algorithms, and design assumptions. Such work does not typically provide direct guidance for a specific artifact, but instead structures how the field analyzes and compares visualization problems and solutions. Brehmer and Munzner's multi-level typology of abstract visualization tasks, for example, provides a portable vocabulary for linking why, how, and what across task descriptions and is explicitly intended to support both design and evaluation \cite{brehmer_multi-level_2013}. Other contributions aim to classify and organize interactions, tasks, and activities \cite{Amar2004, Gotz2008, Lam2008}. Chi's data state reference model similarly frames visualization techniques in terms of abstract data stages and transformation operators, helping make the design space more analyzable and reusable \cite{chi_taxonomy_2000}. Tory and M\"oller likewise position taxonomies as a way to rethink the field at a higher level, proposing a model-based classification intended not only to organize existing approaches but also to guide future research \cite{Tory2004}. These contributions exemplify a distinct and influential form of formalized knowledge in visualization: knowledge that stabilizes the conceptual vocabulary of the field.

A third important form consists of framework-level and theoretical contributions that seek to explain visualization at a more general level. Such efforts are especially notable because visualization has often been described as theoretically sparse, with several authors calling for stronger underlying frameworks and clearer directions for theoretical development \cite{johnson_top_2004, kerren_theoretical_2008, Chen2016}. These contributions include attempts to ground visualization in broader theoretical systems such as information theory \cite{chen_information-theoretic_2010}, distributed cognition \cite{Liu2008}, and scientific modeling \cite{kerren_theoretical_2008}, as well as more explicit reflections on how visualization theory might develop \cite{Chen2016}. Such work does not typically provide direct guidance for producing a particular visualization. Instead, it offers higher-level explanatory, descriptive, or predictive structures for reasoning about the field itself, its central concepts, and the kinds of laws, models, and frameworks that might underpin visualization research more broadly. In the context of design knowledge, these contributions are important not because they prescribe particular design moves, but because they shape how the field conceptualizes what visualization is, how it works, and what kinds of knowledge may count as foundational. At the same time, many of these frameworks are borrowed from neighboring fields rather than developed specifically for visualization design, and their implications for design knowledge in visualization remain unevenly articulated (cf.\ Rogers' discussion of HCI theory borrowing as piecemeal \cite{Rogers2012}).

A fourth important form is process- and model-oriented knowledge. Visualization researchers have developed structured accounts of design activity, including the Nested Model \cite{munzner_nested_2009} and its blocks and guidelines extension \cite{meyer_nested_2015}, the Design Study Methodology \cite{sedlmair_design_2012}, and the Design Activity Framework \cite{mckenna_design_2014}, all of which aim to make the design process more explicit, teachable, and actionable. Other work extends this concern with process into more explicitly optimization-oriented territory \cite{chen_what_2016}. Together, these models and frameworks codify important assumptions about how design proceeds, what counts as a relevant decision, and how process can be structured or improved. They therefore contribute not only local advice, but broader procedural and epistemic models of design work.

A fifth and increasingly consequential form is formalized, computationally operationalized knowledge. Mackinlay's APT is an early and foundational example, explicitly framing automatic chart design as a problem of codifying graphic design criteria so that a presentation tool can systematically generate effective graphical presentations \cite{Mackinlay1986}. Contemporary systems extend this trajectory. Draco and related work represent visualization design knowledge as logical facts, constraints, and preferences over visualization specifications \cite{schmidt_visual_2024}, while Dziban illustrates a complementary concern of how to balance automated recommendation with user agency and prior design intent in the face of ambiguous specifications \cite{lin_dziban_2020}. Recent work has further sought to augment design knowledge bases from empirical data and to probe whether large language models have internalized useful visualization design preferences \cite{kim_data_2026, wang_dracogpt_2025}. Related survey work on automatic visualization and recommendation further underscores how strongly this region of the landscape has developed around codified rules, learned models, and hybrid systems for generating or ranking visualization alternatives \cite{zhu_survey_2020}. In these approaches, design knowledge is not only communicated to human readers but encoded in forms that support recommendation, ranking, generation, and automated completion.

These contributions show that visualization research has been especially successful at building forms of design knowledge that are explicit, portable, and amenable to formalization. This is one of the field's major strengths. Such knowledge supports cumulative scholarship, pedagogy, conceptual stabilization, design tooling, and increasingly powerful forms of computational assistance. At the same time, this region of the landscape privileges what can be stated, abstracted, classified, and encoded with relative stability. It does not encompass the full range of design knowledge needed in practice. In particular, many forms of guidance that are central to design work, including guidelines, heuristics, patterns, methods, and tools, often function less as universal rules than as \emph{intermediate-level knowledge}---resources that orient action across situations without fully determining outcomes.

\subsection{Intermediate-Level Knowledge}

Between universal theory and ultimate particulars lies a broad middle terrain of \emph{intermediate-level knowledge}. In L\"owgren's account, this region includes forms such as design guidelines, usability heuristics, patterns, methods and tools, strong concepts, experiential qualities, and annotated portfolios \cite{Lowgren2013, hook_strong_2012}. These forms are neither fully abstract theories nor merely singular artifacts. Rather, they are abstractions that remain closely tied to design practice, offering resources that can orient action across situations without fully determining outcomes. Intermediate-level knowledge is therefore often generative rather than prescriptive, as it helps designers see possibilities, organize decisions, and navigate recurring challenges while leaving room for adaptation, tradeoffs, and judgment. This view also aligns with broader HCI arguments for generative forms of theory that can bridge explanation and design by offering concepts and principles that remain usable in the construction of new artifacts \cite{beaudouin-lafon_generative_2021}.

The terms gathered here—such as guidelines, heuristics, patterns, methods, and tools—are not introduced as a new typology of subcategories. Rather, they reflect forms of knowledge already named in the design and visualization literature, which often have overlapping and uneven usage.

This framing is useful for data visualization because much of what the field has developed as design guidance appears to inhabit precisely this middle terrain. Consider guidelines, heuristics, principles, and best practices. Although these forms are often grouped together, they do not all function in the same way. Laws and highly general principles tend to sit closer to universal theory, while best practices and guidelines are usually more practice-facing and more closely tied to accumulated precedent, empirical findings, and professional experience. Recent work has shown that visualization guidelines are dispersed across many sources, vary in specificity and grounding, and often require interpretation in context \cite{choi_toward_2021, kim_understanding_2026, ottley_consensus_2026}. Visualization guidance often operates through practice-oriented abstractions that help structure design action while still leaving substantial room for judgment. What unites these forms is not that they all prescribe action with equal force, but that they make design knowledge usable across cases without fully determining what should be done in any one case.

Patterns occupy a similarly important place in this region. In design theory, patterns are a canonical form of intermediate-level knowledge---stylized abstractions that distill key ideas from artifacts or families of artifacts and communicate reusable solutions or design moves \cite{Lowgren2013}. Visualization research has developed several pattern-oriented contributions that fit this description. Bach et al.'s \cite{bach_dashboard_2023} dashboard patterns, for example, are explicitly intended as structured guidance that provides terminology, examples, and ideas for both novice and expert designers, while remaining descriptive rather than prescriptive. Likewise, Deng et al. \cite{deng_revisiting_2023} revisit design patterns of composite visualizations in order to provide a holistic design space and concrete examples that can support reuse, adaptation, and future design. These works exemplify how pattern collections can bridge abstract guidance and concrete practice: they do not specify what the right design must be, but they offer reusable structures for thinking and acting in design situations.

Intermediate-level knowledge also includes methods and tools. L\"owgren explicitly treats design methods and tools as established examples of intermediate-level knowledge, even though they speak to the \emph{how} rather than the \emph{what} of design \cite{Lowgren2013}. This is also visible in visualization. Methods such as the Five Design-Sheet approach aim to scaffold design exploration by giving practitioners a repeatable way to generate, compare, and refine alternatives \cite{roberts_sketching_2016}. Community resources such as VisGuides similarly circulate design-relevant guidance in a practice-facing form that is neither universal theory nor isolated case knowledge \cite{diehl_visguides_2018}. These resources are often most useful not because they dictate specific outcomes, but because they expand designers' capabilities, vocabularies, and repertoires for acting across families of situations.

The intermediate-level framing also accommodates forms of design knowledge that have received less explicit attention in visualization. L\"owgren identifies strong concepts, experiential qualities, and annotated portfolios as additional members of this middle territory \cite{hook_strong_2012, Lowgren2013}. Strong concepts are generative design ideas with demonstrable capacity to inspire subsequent design; experiential qualities concern abstractions related to users' felt experiences rather than only artifact structures; annotated portfolios add knowledge value by relating multiple artifacts through interpretive annotations that reveal more abstract constructs across them \cite{hook_strong_2012, Lowgren2013, zimmerman_designing_2009}. Related HCI work further shows that such forms can serve as bridges between frameworks and concrete design practice, while also noting that these links may become difficult to trace once ideas are absorbed into ongoing design work \cite{law_tracing_2014}. These forms are less established in visualization than guidelines or patterns, but they are potentially valuable precisely because they offer ways of articulating design knowledge that remain grounded in artifacts and practice without collapsing into either isolated examples or universal rules.

Research on practitioner engagement with intermediate-level knowledge underscores the relevance of this category. Gray and Kou \cite{gray_ux_2017} argue that practitioners use and value action-oriented, concrete, and pragmatic forms of intermediate-level knowledge, often intertwined with exemplars and annotated visual material. Their findings suggest that intermediate-level knowledge is not only a theoretical construct for academic discourse, but also a practical resource that helps designers bridge research, precedent, and situated action. Law et al. \cite{law_tracing_2014} similarly show that relations between conceptual frameworks and design practice are often mediated through strong concepts, examples, and design instances rather than through direct top-down application, and that the trace of such knowledge may fade as it becomes assimilated into practice.

Viewed through this lens, intermediate-level knowledge is one of the most important and underarticulated regions of design knowledge in data visualization. It includes many familiar contribution types, but reframes them as a distinct epistemic region rather than as diluted versions of theory or incomplete attempts at formalization. This also suggests why intermediate-level knowledge is so important for both pedagogy and computational support. It is often where design knowledge becomes most usable, but also where ambiguity, interpretation, and adaptation remain indispensable \cite{beaudouin-lafon_generative_2021}.

\subsection{Precedents and Repertoires}

Design knowledge is not carried only in theories, models, or intermediate-level constructs. It is also carried in concrete examples, exemplars, and prior artifacts that help designers recognize possibilities, compare alternatives, and orient judgment. In design theory, such resources are often discussed in terms of \emph{precedents} or \emph{repertoires}: bodies of prior work that can be recalled, consulted, adapted, or recombined in the course of designing \cite{lawson_schemata_2004, boling_nature_2021}. In visualization, these forms of knowledge have received less explicit theoretical attention than guidelines, process models, or taxonomies. Nonetheless, they are clearly present in the field's discourse and practice, where well-known examples, galleries, surveys, and curated collections often function as externalized repertoires, even when their status as authoritative or exemplary remains socially negotiated \cite{baigelenov_how_2025, baigelenov_talking_2026, bako_unveiling_2025, bako_understanding_2023}.

One visible manifestation of this is the role of widely recognized exemplars. Visualization may not have a formal canon in the way some design disciplines do, but it does have recurring reference cases that function in canon-like ways. Historical works such as Minard's depiction of Napoleon's Russian campaign and Nightingale's rose diagrams are repeatedly invoked as enduring examples of expressive, rhetorically powerful, or innovative visualization design \cite{friendly_visions_2002, brasseur_florence_2005}. Their role is not merely historical, however; they are mobilized pedagogically, rhetorically, and evaluatively as touchstones for what visualization can achieve. At the same time, their status as exemplary is not fixed or neutral. Such works are elevated through repeated circulation, retelling, and interpretation, and their meaning as precedents may shift across communities and over time. In this sense, they operate not as uncontested models of ``good'' visualization, but as influential reference points that continue to shape standards, imagination, and judgment long after their original contexts.

Recent visualization research has also begun to examine examples more directly as design resources. Bako et al. \cite{bako_understanding_2023} show that visualization designers actively search for examples in galleries, blogs, portfolios, search engines, published media, and community sites, and use them not only for inspiration but also for understanding data domains, target audiences, implementation strategies, and possible design directions. Their findings suggest that examples function as more than decorative inspiration. Designers use them to generate ideas, validate decisions, refine aesthetics, and translate abstract intentions into concrete design moves. More recent work further shows that the timing, diversity, quantity, and similarity of examples can shape design outcomes, including the number and variety of designs produced and the degree of idea transfer from examples into final work \cite{bako_unveiling_2025}. Related work on inspiration in visualization likewise argues that designers regularly encounter problems that cannot be sufficiently guided by established frameworks alone, and therefore draw on references, influences, and precedent artifacts as part of their design knowledge \cite{baigelenov_how_2025}. These findings suggest an important distinction: precedents may exist as relatively stable external resources, but their design value depends on how they are interpreted, selected, and adapted in relation to a particular situation. In this sense, precedent knowledge is not merely stored in examples; it becomes design-relevant through situated interpretation and use. More broadly, inspiration in visualization extends beyond isolated examples to include wider forms of precedent knowledge, reference, and exemplarity that circulate through professional discourse and practice \cite{baigelenov_talking_2026}.

A related and increasingly important form of precedent knowledge appears in example galleries and curated collections. Lee et al.  \cite{lee_designing_2010} argue that examples offer contextualized instances in which form and content are integrated, and show that users benefit from tools that support browsing and borrowing from multiple examples rather than selecting a single template. More recent work on visualization example galleries likewise shows that galleries are multi-purpose structures that can function as documentation, tutorial resources, marketing material, reusable template corpora, and even test infrastructure \cite{yang_considering_2024}. These findings suggest that galleries do not merely display finished work; rather, they externalize repertoires. They make possible forms, patterns, and implementation strategies available for browsing, comparison, and reuse, thereby turning otherwise private or tacit repertoires into more collective design resources.

This perspective also sheds light on the role of surveys, catalogs, and broad collections of visualization examples. Although such contributions are often framed as literature reviews, corpora, or design-space overviews, they can also function as formalized repertoires. By organizing and displaying families of techniques, examples, or cases, they provide designers and researchers with an external reference structure for recalling possibilities, comparing alternatives, and identifying promising directions. Similarly, systems such as Many Eyes and Tableau Public created shared public spaces where users could upload data, generate visualizations, and discuss them, effectively broadening the visibility and circulation of examples at community scale \cite{viegas_manyeyes_2007, morton_public_2014}. These resources do not replace theory or intermediate-level abstractions, but they provide a different kind of design knowledge---one grounded in concrete prior works and the design possibilities they make available.

Viewed through this lens, precedents and repertoires form an important region of design knowledge in visualization. They are more concrete than intermediate-level abstractions such as guidelines or patterns, yet more explicit and shareable than wholly tacit knowledge. At present, this region appears influential in practice but comparatively underarticulated in scholarship. Recognizing it more explicitly has several benefits. It clarifies why designers so often work through examples, why galleries and curated collections matter, and why concrete artifacts can carry knowledge that is not reducible to general rules. It also suggests opportunities for future work, including annotated precedent collections, richer gallery infrastructures, and stronger support for case-based reasoning in both design pedagogy and computational tools. At the same time, repertoires are not neutral collections. What becomes exemplary, memorable, or reusable is shaped by circulation, visibility, and the evaluative norms of the communities in which examples are encountered and discussed \cite{baigelenov_talking_2026}. Even widely recognized precedents should therefore be understood not as self-evidently canonical but as artifacts whose authority is continually reproduced and negotiated in discourse and practice.

\begin{figure}[h]
    \centering
    \includegraphics[width=1\linewidth]{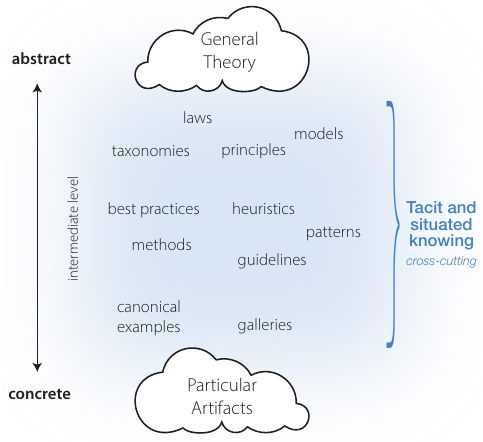}
    \caption{A simplified view of the design-theoretical terrain between general theory and particular artifacts. Intermediate-level knowledge occupies the broad middle region, including forms such as best practices, methods, guidelines, heuristics, and patterns. In visualization, many familiar forms of design guidance appear to inhabit this intermediate space, positioned between highly abstract theoretical constructs and concrete examples or artifacts. Adapted from L\"owgren \cite{Lowgren2013}.}
    \label{fig:intermediate-level}
\end{figure}

Figure~\ref{fig:intermediate-level} provides a simplified illustration of the terrain between general theory and particular artifacts. The figure highlights intermediate-level knowledge as a broad middle region rather than a narrow category, populated by forms such as best practices, methods, guidelines, heuristics, and patterns. This framing is useful for visualization because many of the field's most familiar forms of design guidance are neither universal theories nor singular examples, but abstractions that orient action across situations without fully determining outcomes.

\subsection{Tacit and Situated Knowing}

At the most practice-proximal end of the epistemic landscape are forms of design knowledge that are difficult to fully articulate, detach from context, or stabilize in explicit form. These include tacit knowing, situated judgment, contextual interpretation, and the kinds of repertoires that become embodied through experience rather than only externalized in rules, models, or examples. Unlike the preceding regions, tacit and situated knowing is not only a distinct region of the landscape, but also a cross-cutting aspect of how other forms of design knowledge become meaningful in practice.

In design theory and related philosophical work on practice, such forms of knowing are central because good action does not proceed only through the application of abstract knowledge to predefined problems. It also involves responding to unique situations, framing what matters, interpreting constraints, and making judgments shaped by context, values, and experience \cite{schon_reflective_1983, buchanan_wicked_1992, nelson_design_2012, dunne_back_1997, polanyi_personal_1958}. In this view, situated knowing is not simply what remains after formalization fails; it arises because abstraction necessarily leaves some particulars unspecified, while action must still succeed in the concrete case.

Our concern here is not with all tacit knowledge that may become relevant in a visualization project. Rather, we refer to tacit forms of knowing insofar as they enter into visualization design activity itself---for example, in judging relevance, interpreting constraints, framing what matters, adapting guidance, or determining what is appropriate in a concrete situation. In practice, these moments often draw on both design-specific and domain- or situation-specific understandings. The concern in this paper is with tacit knowing as it becomes operative in design judgment and action.

Historically, this region has received less explicit attention in visualization than more formal and codified forms of knowledge. Much of the field's design discourse has emphasized empirical findings, taxonomies, models, process frameworks, and increasingly computationally operationalized knowledge. These contributions are important, but they do not fully capture the more situated and interpretive dimensions of design practice. Recent work in visualization has begun to acknowledge this limitation more directly. Meyer and Dykes \cite{meyer_criteria_2020}, for example, argue for a broader epistemological perspective in visualization design studies, emphasizing that design knowledge may be multiple, situated, and socially constructed rather than solely objective, universal, or formalizable. Related work has likewise begun to foreground alternative epistemological perspectives in visualization, including feminist accounts that stress entanglement, positionality, and the conditions through which visualization knowledge is produced \cite{akbaba_entanglements_2025}. These contributions open important space for recognizing forms of knowledge that are inseparable from local practice, interpretation, and social context.

Recent empirical work in visualization design practice has also begun to make these forms of knowing more visible. Parsons et al. \cite{parsons_design_2020} argue that design judgment in visualization practice is layered, situated, and personal, and that existing literature has historically emphasized more formal and codifiable forms of knowledge while paying less attention to how practitioners navigate ambiguity, context, and evolving design situations. Related work on framing and problem--solution co-evolution further shows that visualization designers do not simply apply expertise within a predefined design space; they actively construct and revise that space through framing, visual experimentation, and iterative judgment \cite{parsons_beyond_2026}. Complementing these project-level accounts, recent episode-level research shows that practitioners make unresolved visualization work actionable through provisional moves that reveal interpretive possibilities, clarify tractability, and sometimes reorient the work \cite{parsons_situatedness_2027}. Practitioner accounts also suggest that codified knowledge may be internalized rather than discarded---theories, principles, and guidelines may not be explicitly referenced, but instead operate as backgrounded repertoire, habit, or judgment in action. In this view, important aspects of design knowledge are not only stored in explicit artifacts such as theories, patterns, or precedents, but enacted through situated engagement with the problem at hand.

Recent studies of visualization work in context reinforce this point. Zhang et al.'s \cite{zhang_visualization_2023} study of COVID-19 dashboard creators shows that visualization design unfolds through shifting sociotechnical circumstances, public demands, maintenance burdens, and changing phases of work, from initial creation to expansion, upkeep, and eventual termination. Dhawka and Dasgupta \cite{dhawka_social_2025} similarly show that practitioners understand visualization work as shaped by beliefs, values, biases, politics, and questions of neutrality, explicitly framing visualization production as socially constructed and grounded in situated knowledges. These studies broaden the meaning of situated knowing in visualization. It is not only a matter of individual tacit skill, but also of how designers operate within organizational, political, relational, and ethical conditions that shape what can be made and how it is judged.

This perspective is also visible in recent studies of inspiration and example use. Designers do not treat examples only as static references; they interpret, adapt, combine, and sometimes resist them in relation to particular goals, constraints, and values \cite{bako_understanding_2023, bako_unveiling_2025, baigelenov_how_2025, baigelenov_talking_2026}. What matters in such moments is often not simply access to a body of explicit knowledge, but the capacity to judge relevance, see latent possibilities, and recognize what is appropriate in context. These are hallmarks of tacit and situated knowing: forms of knowledge that may draw on codified guidance and prior examples, but are not reducible to them.

Viewed through this lens, tacit and situated knowing is not a residual category containing whatever cannot be formalized. It is an essential part of how design knowledge operates in practice. It is where rules meet exceptions, where precedents are interpreted rather than copied, and where design moves are judged in relation to concrete circumstances. To borrow Dunne's formulation, this is the ``rough ground'' of practice, where technique alone is insufficient and practical judgment remains indispensable \cite{dunne_back_1997}. Polanyi's account of tacit knowing sharpens the point further, as he argues that competent action depends on personal, skillful integrations that cannot be exhaustively specified without losing what makes them effective in the first place \cite{polanyi_personal_1958}. This region of the landscape remains comparatively underdeveloped in visualization scholarship, but it is increasingly visible in recent work on design practice, judgment, framing, epistemology, and socially situated visualization work. Recognizing it more explicitly is important not only for understanding how visualization design actually works, but also for clarifying the limits of purely codified accounts of expertise and the kinds of human involvement that remain central in AI-assisted design.

\begin{table*}[t]
\centering
\footnotesize
\setlength{\tabcolsep}{3pt}
\renewcommand{\arraystretch}{1.4}

\begin{tabularx}{\textwidth}{
>{\raggedright\arraybackslash}p{0.12\textwidth}
>{\raggedright\arraybackslash}p{0.18\textwidth}
>{\raggedright\arraybackslash}p{0.17\textwidth}
>{\raggedright\arraybackslash}X
>{\raggedright\arraybackslash}X
}
\hline
\textbf{Region} &
\textbf{Forms} &
\textbf{Affords} &
\textbf{Representative VIS Contributions} &
\textbf{Limits / Risks} \\
\hline

\textbf{Formal, abstract, codified knowledge}
&
Empirical findings; taxonomies and typologies; theoretical frameworks; process models; knowledge bases; formal constraints
&
Portability; cumulative structure; explicitness; shared vocabulary; formalization
&
Perceptual and cognitive principles; task typologies; Nested Model and Design Study Methodology; Draco and related knowledge bases; optimization-oriented recommender frameworks
&
Can overprivilege what is explicit, stable, and generalizable; may underrepresent contextual interpretation and local contingencies \\ \addlinespace[4pt]

\textbf{Intermediate-level knowledge}
&
Guidelines; heuristics; patterns; methods and tools; strong concepts; experiential qualities; annotated portfolios
&
Generative guidance across situations; actionable abstraction; bridge between theory and practice
&
Visualization guideline collections; dashboard and composite visualization patterns; Five Design-Sheet; VisGuides; design resources
&
Can be unevenly grounded, open to interpretation, or difficult to validate in uniform ways; may fade into practice and become hard to trace \\ \addlinespace[4pt]

\textbf{Precedents and repertoires}
&
Exemplars; prior artifacts; galleries; curated collections; surveys as repertoires; public repositories
&
Example-based orientation; comparison; adaptation; repertoire building; case-based reasoning
&
Minard and Nightingale as recurring exemplars; D3 and Vega example galleries; Many Eyes; Tableau Public; curated corpora and surveys
&
May remain informal, fragmented, or unevenly theorized; what becomes visible as precedent is shaped by circulation and community norms \\ \addlinespace[4pt]

\textbf{Tacit and situated knowing}
&
Tacit skill; situated judgment; contextual interpretation; framing in practice; embodied repertoires; socially situated knowing
&
Fit to particular situations; handling ambiguity; weighing competing constraints; sustaining judgment in context
&
Vis practice studies; studies of design judgment and decision making; socially situated studies of visualization work
&
Difficult to formalize, externalize, teach, or validate through conventional publication genres alone; easily backgrounded \\

\hline
\end{tabularx}

\caption{Summary of the major regions of the design knowledge landscape in data visualization. The regions are analytically distinguishable, but in practice they often interact and overlap through knowledge-in-use.}
\label{tab:design-knowledge-landscape}
\end{table*}

\section{Reconsidering Design Knowledge in Visualization}

The epistemic landscape outlined above makes it possible to reconsider how design knowledge has developed in visualization research. The goal is not to rank forms of knowledge or treat some as more legitimate than others, but to clarify where the field has built especially strong knowledge infrastructures, where important regions remain more fragmented or weakly articulated, and why design expertise depends not only on knowledge artifacts themselves but on how they are enacted in practice. Viewed this way, the landscape is both descriptive and diagnostic, as it reveals the diversity of design knowledge already present in visualization while also showing where the field's ways of naming, studying, and supporting that knowledge remain uneven.

\subsection{Strong Knowledge Infrastructures}

One of the clearest patterns in the landscape is that visualization research has built especially strong infrastructures around forms of knowledge that are explicit, portable, and reusable. These include formalized empirical findings, taxonomies and typologies, process and design-study models, and increasingly computationally operationalized representations of design knowledge \cite{brehmer_multi-level_2013, munzner_nested_2009, sedlmair_design_2012, schmidt_visual_2024, kim_data_2026, moritz_formalizing_2019}. Together, these contributions have given the field durable ways of accumulating results, stabilizing terminology, teaching design, and embedding knowledge in tools and systems.

Intermediate-level knowledge is also well represented in some areas, especially through guidelines, heuristics, patterns, methods, and design resources that help practitioners and researchers navigate recurring problems \cite{Heer2006, choi_toward_2021, bach_dashboard_2023, deng_revisiting_2023, roberts_sketching_2016, Elmqvist2011, Sedig2013}. Although these forms are often less formal than theories or knowledge bases, they still benefit from being codified and shared in recognizable formats. This helps explain why guidelines, pattern collections, and structured design methods have become such visible and influential contributions. They occupy an especially productive middle ground: close enough to practice to be usable, yet abstract enough to travel across cases.

Taken together, these strong knowledge infrastructures are one of the field's major accomplishments. They make it possible to build cumulative scholarship, translate research into reusable resources, and increasingly connect design knowledge to computational support \cite{schmidt_visual_2024, kim_data_2026, wang_dracogpt_2025}. At the same time, their very strengths help reveal an asymmetry in the landscape. The forms of design knowledge that visualization has stabilized most successfully tend to be those that can be abstracted, codified, organized, and circulated with relative clarity.

\subsection{Fragmented and Underarticulated Regions}

Other regions are not absent so much as fragmented, emergent, or weakly integrated. This is especially true for precedents and repertoires, tacit and situated knowing, and some intermediate-level forms such as strong concepts, experiential qualities, and annotated portfolios. Visualization clearly has examples, exemplars, galleries, and recurring reference cases that function as precedent-like resources, and recent work shows that designers rely heavily on them in practice \cite{bako_understanding_2023, bako_unveiling_2025, lee_designing_2010, yang_considering_2024}. Yet the field has fewer explicit ways of theorizing these resources as design knowledge in their own right, or of distinguishing how they function differently from guidelines, patterns, or formal models.

A similar point holds for tacit and situated knowing. Recent visualization scholarship has increasingly acknowledged that design is shaped by interpretation, framing, values, context, and socially situated practice \cite{meyer_criteria_2020, parsons_design_2020, parsons_beyond_2026, dhawka_social_2025, zhang_visualization_2023}. However, compared with the field's well-developed traditions of codified knowledge, these forms remain less systematically studied, named, or supported. They often appear indirectly through empirical accounts of practice, reflections on rigor or epistemology, or studies of example use and inspiration, rather than through explicit frameworks for understanding them as design knowledge.

The result is not a simple absence, but an uneven landscape. Some forms of design knowledge have strong publication genres, established methods of validation, and clear pathways into pedagogy and tools. Others circulate more diffusely through examples, discourse, communities of practice, and lived experience \cite{otto_visualization_2024, baigelenov_talking_2026, dhawka_social_2025}. Recognizing this unevenness matters because it affects what the field finds easy to value, formalize, and support. It also affects what kinds of design expertise become visible to researchers and what kinds remain backgrounded.

\subsection{Knowledge-in-Use}

The landscape also suggests that design knowledge in visualization cannot be understood only as a collection of knowledge artifacts. Theories, guidelines, patterns, precedents, and repertoires are all important, but in practice they rarely operate in isolation. Designers move among them as they frame problems, generate alternatives, evaluate tradeoffs, and judge what is appropriate in context \cite{parsons_design_2020, parsons_beyond_2026, bako_understanding_2023, bako_unveiling_2025}. What matters, then, is not only that these forms of knowledge exist, but how they are interpreted, combined, adapted, and sometimes resisted in use.

This shift from knowledge artifacts to knowledge-in-use is important because it reframes design expertise. Expertise is not simply possession of more knowledge, nor only access to better codified resources. It also involves the situated capacity to mobilize multiple forms of knowing in relation to the demands of a particular case. A guideline may inform a decision, a precedent may suggest a direction, a pattern may provide structure, and tacit judgment may determine whether any of them truly fit the situation at hand \cite{dunne_back_1997, parsons_design_2020, dhawka_social_2025}. Some forms of codified knowledge may also be so thoroughly absorbed into practice that they no longer appear as explicit resources at all, persisting instead as backgrounded repertoire, habit, or judgment-in-action. The significance of a knowledge resource therefore depends not only on its content, but on how it is enacted in concrete situations.

Seen in this way, the epistemic landscape is not merely a typology of knowledge forms. It is also a way of understanding the movement among them. Abstract knowledge becomes design-relevant only through interpretation and application; precedents become useful only when selected and adapted; tacit repertoires are developed through repeated engagement with concrete situations \cite{bako_understanding_2023, zhang_visualization_2023, parsons_beyond_2026}. Knowledge-in-use names this ongoing coordination. It is where codified resources meet the contingencies of practice, where precedents are made meaningful, where abstract guidance is transformed into situated action, and where design outcomes are ultimately shaped. In practice, the boundaries among these forms matter less than the ways designers move among them: from theory to guidance, from examples to judgment, and from repeated action to tacit repertoire.

This has direct consequences for how visualization research conceptualizes design support. If design knowledge is understood only as a body of explicit artifacts, then support may be framed mainly in terms of storage, retrieval, recommendation, or rule application. But if design also depends on knowledge-in-use, then support must also account for how knowledge is made relevant, how judgment is sustained, how different resources are coordinated in context, and how new tools may reshape the work through which design knowledge is enacted. This point is especially important in the implications that follow for research, pedagogy, and AI-assisted design.

\section{Implications and Future Directions}

The epistemic landscape developed in this paper is not only descriptive. It also has implications for what kinds of design knowledge visualization research chooses to build, how such knowledge is documented and circulated, and how design support is conceived in both pedagogy and computational systems. If visualization has developed especially strong infrastructures for codified and portable forms of knowledge, then one challenge is not simply to continue extending those forms, but also to strengthen how the field recognizes, studies, and supports other regions of the landscape. The following sections consider what this broader account suggests for research, education, and AI-assisted design.

\subsection{Research Agenda}

One implication of this framework is the need for a broader research agenda around design knowledge in visualization. Some regions of the landscape are already comparatively well developed, especially formalized empirical findings, taxonomies, process frameworks, and computationally operationalized knowledge \cite{choi_toward_2021, brehmer_multi-level_2013, munzner_nested_2009, sedlmair_design_2012, schmidt_visual_2024, kim_data_2026}. Other regions remain more fragmented or weakly articulated, even when they are clearly influential in practice \cite{bako_understanding_2023, bako_unveiling_2025, parsons_design_2020, parsons_beyond_2026}. This points to several productive directions for future work.

A first direction concerns intermediate-level knowledge beyond familiar forms such as guidelines and patterns. Visualization has developed many resources that already function in this middle terrain, but some forms remain much less established, including strong concepts, experiential qualities, and annotated portfolios. These forms may be especially valuable because they preserve links to practice and artifacts without collapsing into either isolated examples or universal rules. Future work could examine how such contributions might be developed, documented, and evaluated in ways appropriate to visualization as a field.

A second direction concerns precedents and repertoires. The field clearly relies on examples, galleries, surveys, and recurring reference cases \cite{bako_understanding_2023, lee_designing_2010, yang_considering_2024, baigelenov_talking_2026}, yet these resources are still only weakly theorized as design knowledge. This raises questions about how precedent collections might be curated, how galleries might better support comparison and adaptation, and how community repertoires are formed, circulated, and contested. It also suggests that surveys, corpora, and repositories may be doing more epistemic work than is typically acknowledged, functioning not only as reviews or archives but as infrastructures for case-based reasoning and design orientation.

A third direction concerns tacit and situated knowing. Recent work has begun to show that visualization design depends on framing, judgment, interpretation, social context, and the situated conditions under which design knowledge is enacted \cite{meyer_criteria_2020, parsons_design_2020, parsons_beyond_2026, dhawka_social_2025, zhang_visualization_2023}. Yet these dimensions remain difficult to study using the publication genres and evaluative norms that have historically dominated the field. Future work could therefore expand empirical attention to how designers internalize knowledge over time, how codified guidance becomes backgrounded in practice, how values and organizational settings shape design action, and how design expertise develops through lived engagement rather than only through explicit rule acquisition.

A fourth direction concerns how designers move among multiple forms of knowledge in practice. If design expertise depends not only on the existence of knowledge artifacts but on the ability to move among them, then an important research question is how such movement occurs. How do designers decide when to rely on guidelines, when to adapt or ignore them, when to draw on precedents, and when to privilege tacit judgment? How are multiple forms of knowing coordinated during framing, iteration, and evaluation? These questions suggest that future work should not only catalog design knowledge, but also study how it is enacted, transformed, and made relevant in practice \cite{parsons_beyond_2026, bako_understanding_2023, law_tracing_2014}.

These directions suggest that a richer account of design knowledge in visualization will require broader publication forms, empirical methods, and evaluative criteria. The issue is not merely to fill missing categories, but to strengthen the field’s capacity to recognize and support the full range of knowing through which visualization design works.

\subsection{Education and Practice}

This conceptual lens suggests implications for how visualization design is taught and supported in practice. If design knowledge is treated primarily as codified guidance, then education may emphasize principles, rules, and best practices as though competent design were mainly a matter of correct application. Such resources are important, but they are not sufficient. Design education must also prepare students to interpret guidance, build repertoires, work from precedents, and develop judgment in situations where rules do not fully determine action \cite{parsons_preparing_2023}.

This suggests a more expansive pedagogical approach. Rather than treating guidelines as endpoints, educators might treat them as resources to be interpreted alongside examples, competing constraints, and concrete design situations. Galleries, annotated collections, and structured repertoires of examples may therefore be as important pedagogically as principles and taxonomies. Likewise, methods and tools should be understood not only as procedural aids, but as ways of helping students develop capacities for exploration, comparison, and reflection \cite{gray_building_2023}.

A similar point applies to practice. Practitioners rarely work by applying one form of knowledge at a time. They work through combinations of theory, precedent, heuristics, methods, examples, and tacit judgment, often under time pressure, organizational constraints, and shifting problem definitions \cite{parsons_design_2020, bako_understanding_2023, zhang_visualization_2023}. Supporting practice therefore means more than offering better rules. It also means building infrastructures that help designers compare cases, interpret guidance in context, and develop repertoires that can be mobilized flexibly across situations.

More broadly, different regions of the landscape are likely to be emphasized differently across settings. Education may foreground explicit guidance and repertoire-building, while practice may rely more heavily on precedent, adaptation, and judgment in context. One reason the research--practice gap persists may be that the field has often assumed that knowledge transfer should proceed mainly through explicit guidance or generalizable findings. But if much design knowledge operates through intermediate forms, precedents, and situated enactment, then closing the gap may also require richer case-based resources, more practice-facing archives, and stronger attention to how knowledge becomes usable in context rather than merely available in principle \cite{kim_understanding_2026, otto_visualization_2024}.

\subsection{AI-Assisted Design}

The epistemic landscape developed here also clarifies what is at stake in AI-assisted design. One implication is that computational support is not an all-or-nothing matter. Some forms of design knowledge are more readily formalized, retrieved, and operationalized than others. Formalized empirical findings, taxonomies, patterns, and design constraints are comparatively amenable to knowledge bases, recommendation systems, and LLM-based retrieval and prompting \cite{moritz_formalizing_2019, schmidt_visual_2024, kim_data_2026, wang_dracogpt_2025}. By contrast, precedents, tacit repertoires, situated judgment, and problem framing are more difficult to represent in stable and portable ways. The implications of AI-supported design are likewise likely to vary by context, since systems intended for novice guidance, professional practice, or infrastructure-building may draw on different regions of the landscape and should not be expected to support design knowledge in the same way.

Additionally, the issue is not simply one of dividing labor between humans and computational systems. As work on human--automation coordination has long argued, automation changes work qualitatively rather than merely reallocating fixed functions \cite{dekker_maba-maba_2002, hollnagel_role_2003}. In the context of visualization design, formalizing some forms of knowledge for AI support may reshape the timing, visibility, and coordination of design activity itself. Some forms of work may be foregrounded, others backgrounded; some judgments may be scaffolded, others displaced or deferred \cite{parsons_myths_2026, woods_anticipating_2000}. The relevant question is therefore not only what AI systems can generate or retrieve, but how they reorganize the conditions under which design knowledge is enacted.

This is where the distinction between knowledge artifacts and knowledge-in-use becomes especially important. Systems may be highly effective at surfacing codified guidance, summarizing alternatives, or recombining known structures, yet still remain limited in supporting the situated work of framing, selecting what matters, and judging what is appropriate in context. Indeed, one risk of AI-assisted design is that design work may be implicitly redefined around the forms of knowledge that are easiest to formalize, thereby obscuring the continuing importance of tacit, precedent-based, and situationally embedded forms of knowing \cite{kim_how_2025}.

A more productive approach is to treat AI as support for particular regions of the landscape rather than as a substitute for design expertise as a whole. This suggests not only differentiated expectations for AI systems, but also differentiated criteria for evaluating them. The question is not simply whether a system produces plausible outputs, but which forms of design knowledge it draws on, which it neglects, and how its use reshapes the work of design over time. From this perspective, the future of AI-assisted visualization design depends less on replacing human judgment than on understanding how computational systems can participate responsibly in a broader ecology of design knowing.

\section{Conclusion}

Data visualization research has developed a substantial body of design knowledge, including empirical findings, guidelines, patterns, process models, taxonomies, and increasingly formalized computational representations. In this paper, we have argued that these contributions are best understood as part of a broader epistemic landscape spanning formal and highly codified knowledge, intermediate-level knowledge, precedents and repertoires, and tacit and situated knowing.

Using this landscape as a conceptual lens, we have shown that visualization research has built especially strong infrastructures for codified and portable forms of knowledge, while other regions remain more fragmented or less explicitly articulated. We have further argued that visualization design depends not only on knowledge artifacts, but also on \emph{knowledge-in-use}: the situated interpretation, adaptation, and coordination of multiple forms of knowing in practice.

Our aim has not been to diminish existing forms of visualization knowledge, but to situate them within a broader account of how design knowledge works in the field. Recognizing this broader landscape helps clarify what kinds of knowledge visualization research has developed most strongly, what remains underarticulated, and why future work on design, pedagogy, and AI-assisted systems must attend not only to knowledge artifacts, but also to how they are enacted in practice.






\acknowledgments{%
  This work was supported in part by a grant from the NSF (\# 2146228).%
}

\bibliographystyle{abbrv-doi-hyperref}

\bibliography{references}

\end{document}